\documentstyle[12pt]{article}
\newcommand{\bra}{\langle}
\newcommand{\ket}{\rangle}
\newcommand{\sigz}{\Sigma^0}
\newcommand{\lamz}{\Lambda^0}
\newcommand{\sigs}{\Sigma^*}
\newcommand{\lams}{\Lambda^*}
\newcommand{\lama}{\Lambda(1520)}
\newcommand{\lamb}{\Lambda(1405)}

\newcommand{\glla}{\Gamma(\Lambda(1520) \rightarrow \Lambda^0 + \gamma)}
\newcommand{\gllb}{\Gamma(\Lambda(1405) \rightarrow \Lambda^0 + \gamma)}
\newcommand{\glsa}{\Gamma(\Lambda(1520) \rightarrow \Sigma^0 + \gamma)}
\newcommand{\glsb}{\Gamma(\Lambda(1405) \rightarrow \Sigma^0 + \gamma)}
\newcommand{\cl}{\centerline}
\newcommand{\beq}{\begin{equation}}
\newcommand{\eeq}{\end{equation}}
\newcommand{\beqa}{\begin{eqnarray}}
\newcommand{\eeqa}{\end{eqnarray}}

\begin{document}

\thispagestyle{empty}
\setcounter{footnote}{2}
\renewcommand{\thefootnote}{\fnsymbol{footnote}}
\hfill {\tt SUNYSB-NTG-92-07}

\hfill {\tt September 1992}
\vskip 0.25in
\cl{\Large\bf{Electromagnetic Decays of Excited Hyperons. (II)}\footnote{This
work is supported in parts by grants from the NSF and the U.S. Department of
Energy under contract \#DE-FG02-88ER40388.}}
\vskip 0.3in
\cl{Yasuo Umino\footnote{Address after September 1, 1992: NIKHEF-K, Postbus
41882, 1009 DB Amsterdam, The Netherlands.}}
\vskip 0.1in
\cl{\it{Department of Physics}}
\cl{\it{State University of New York at Stony Brook}}
\cl{\it{ Stony Brook, NY 11794, U.S.A.}}
\vskip 0.075in
\cl{and}
\vskip 0.075in
\cl{Fred Myhrer}
\vskip 0.10in
\cl{\it{Department of Physics and Astronomy}}
\cl{\it{University of South Carolina}}
\cl{\it{Columbia, SC 29208, U.S.A.}}
\vskip 0.3in
\cl{\large\bf{Abstract}}
\vskip 0.2in
Excited negative parity hyperon masses are calculated in a chiral
bag model in which the pion and the kaon fields are treated as perturbations.
We also calculate the hadronic  widths of $\lama$ and $\lamb$ as well as the
coupling
constants of the lightest $I=0$ excited hyperon to the meson-baryon channels,
and discuss how the dispersive effects of the hadronic
meson-baryon decay channels affect the excited hyperon masses.
Meson cloud corrections to the electromagnetic decay widths of the two
lightest excited hyperons into ground states $\lamz$ and $\sigz$
are calculated within the same model and are found to be small.
Our results strengthen the argument that
predictions of these hyperon radiative decay widths provide an excellent test
for various quark models of hadrons.
\vskip 0.3in
\cl{Submitted to: {\it Nuclear Physics A}}
\vfill\eject
%
%
\setcounter{footnote}{0}
\renewcommand{\thefootnote}{\arabic{footnote}}
\baselineskip = 18pt
\section{Introduction}

With the anticipated completion of CEBAF and the proposed KAON laboratory,
there has been a renewed interest in the study of low-lying excited
hyperons.$^{\cite{bar}}$
Although the existence of these hyperon states has been established for quite
some time,$^{\cite{pil}}$ their underlying quark structure is not yet well
understood.
One reason is our poor knowledge of the excited hyperon mass
spectrum$^{\cite{pdg}}$ which introduces uncertainties when
building models of hadrons. Another reason is that the measurements of
transition amplitudes involving excited hyperon states, which test model
dependent wavefunctions and transition operators, have not been made
with sufficient accuracy.
Theoretically, a relatively simple type of such transitions is the
electromagnetic one since corresponding transition
currents and thus, the operators, can, in principle, be constructed
once the model Lagrangian and assumptions about the hadron structure are
specified.
Fortunately, calculations of electromagnetic transition amplitudes between
excited and ground state hyperons are found to be strongly model
dependent$^{\cite{dhk,wpr,kms,um2}}$ and their
measurements, which are being planned at CEBAF,$^{\cite{cebaf}}$
may help to determine the hyperon wave function composition and thereby put
strong restrictions on possible phenomenological models of hadrons.

Hyperon resonances also play an important role in low-energy $\bar{K}N$
interactions which are
characterized by the presence of coupled two-particle channels.
The $\pi\Sigma$ channel can couple to the low-energy $\bar{K}N$ system with
isospin $I=0$ whereas both the $\pi\Sigma$ and $\pi\Lambda$ channels are open
when the $\bar{K}N$ system is coupled to $I=1$.
The lightest hyperon resonance, $\lamb$, plays a crucial role in
understanding the properties of the $K^-$ atoms and to determine the structure
of this state is of prime importance in $K^-$ atomic studies.
It has long been speculated that the
$\lamb$, being close to the $\bar{K}N$ threshold, is a
candidate for the $K^-p$ bound state interpretation,$^{\cite{cbag,sw,lan}}$
however, a similar analysis has not been made for $\Lambda(1520)$.
A Cloudy Bag Model analysis (to be discussed later)
of low-energy S-wave $\bar{K}N$ scattering shows
that $\lamb$ is mostly a meson-baryon bound state.$^{\cite{cbag}}$
Also, in the bound-state approach to the hyperons in the Skyrme
model,$^{\cite{chk}}$ $\lamb$ emerges naturally as a bound state of the
strangeness carrying kaon and the $SU(2)$ soliton.
This suggests that $\lamb$ might have a dominant molecular $q^4\bar{q}$
structure instead of being a pure $qqq$-state.
In an effective meson-baryon theory
the mass and the width of $\lamb$ may be reproduced in
a simple coupled channel
$K$-matrix analysis of $\pi\Sigma$ scattering.$^{\cite{pil,dal}}$
For further developments see {\it e.g.} Oades and Rasche$^{\cite{or}}$
and Williams, Ji and Cotanch.$^{\cite{wjc}}$

In contrast, both the NRQM$^{\cite{ik,clo}}$ and chiral bag model
calculations$^{\cite{um2,um1}}$
find that the {\it lightest} $\Lambda^*(\frac{1}{2}^-)$
$qqq$ state is almost mass degenerate with the {\it lightest}
$\Lambda^*(\frac{3}{2}^-)$ $qqq$ state.
In addition, the
non-relativistic quark model (NRQM) finds
$\lamb$ to be dominantly a three quark
($qqq$) flavor singlet state$^{\cite{ik,clo}}$ whereas in bag models it
is about an
even mixture of flavor singlet and octet states.$^{\cite{um2,deg}}$
The chiral bag model, like the Cloudy Bag Model, has a specific coupling to the
open hadronic channels through the meson cloud.
An improved treatment of this meson cloud
({\it i.e.}, a proper inclusion of dispersive effects of the hadronic widths)
might explain the observed
$\vec{L}\cdot\vec{S}$ splitting of the lowest $J^P=\frac{1}{2}^-$
and $\frac{3}{2}^-$ $\Lambda^*$-states,$^{\cite{tor,ay,sg,bonn}}$
an effect we shall
%
estimate within our model in this work.
If the $\lamb$ is a molecular (or quasi-bound
$\bar{K}N/\pi\Sigma$) state, then a further and so far un-observed low-lying
three-quark $\lams (\frac{1}{2}^-)$ state should exist.
Thus, establishing all the low-lying excited hyperons and
probing the quark substructure of these hyperon resonances through,
for example, their
radiative transition rates will contribute
towards a better understanding of the nature of the $\lamb$ state
and its role in low-energy $\bar{K} N$ interactions.

In two previous articles$^{\cite{um2,um1}}$ we calculated the masses and
wavefunctions of the low-lying negative parity baryon resonances as well as
some hyperon radiative decay widths relevant to a planned CEBAF
experiment$^{\cite{cebaf}}$ using a perturbative version of the chiral bag
model. In calculating the excited baryon masses
we found that both the one-gluon exchange interactions and the
dispersive effects from the pion-baryon channels play important
roles in describing the mass-spectrum.
(See Ref.~\cite{um2} for errata to
Tables~4 and 5 of Ref.~\cite{um1}. Erratum to the figures and other tables of
Ref.~\cite{um1} are available upon request.)

In this paper we extend our work to include the
kaon cloud in our model calculation of the hyperon mass spectrum
and also estimate the effects due to
mass differences between initial/final and
intermediate baryons in the evaluation of baryon self-energy diagrams.
Furthermore, in Ref.~\cite{um2} we calculated only
the quark core contributions to the radiative decay widths
of $\lama$ and $\lamb$ decaying into ground states $\lamz$ and $\sigz$.
These decay widths were found to be much smaller than those calculated in the
NRQM$^{\cite{dhk}}$ due to the different spin-flavor content of excited
hyperons in the two models.
For example, the chiral bag model predicts, similar to the early but incomplete
MIT bag calculations of these states,$^{\cite{deg}}$
a large but non-dominant
admixture of flavor octet component in $\lamb$,
a state which has traditionally been treated as a flavor singlet state based on
the results of NRQM calculations.$^{\cite{ik,clo}}$ This octet admixture
reduces drastically the radiative decay widths of the lightest
$\Lambda^*(\frac{1}{2}^-)$ state.
The small hyperon radiative decay widths found in Ref.~\cite{um2} imply that
meson cloud corrections to the model might affect the calculated widths, and
therefore  we calculate in this work the meson cloud corrections to the above
decays widths.

This paper is organized as follows. In the following section we present the
excited hyperon mass spectrum calculated
with the kaon fields included in the meson cloud. We
discuss how the dispersive effects of our calculated
hadronic widths contribute to the hyperon masses and
examine some of the difficulties encountered in evaluating the
meson cloud contributions to excited baryon masses.
The strong decay widths and coupling constants of $\lama$ and $\lamb$ are
then estimated with our model. In Section 3 we present the meson cloud
corrections to the excited hyperon radiative decay widths. Meson
electromagnetic
transition currents constructed to calculate these corrections are, in
general, two quark operators acting on spin-flavor wavefunctions of a
three-quark system. Finally in Section 4, we conclude with a summary and
a discussion of the results of the present work.

\section{Low-lying Negative Parity Hyperons}

\subsection{The excited hyperon mass spectrum}

In this section we extend our calculation of the low-lying negative parity
hyperon mass spectrum
by including the kaon cloud contribution while
neglecting the effects of the $\eta$ cloud.
These massless meson fields
are determined by a boundary condition on the bag surface
requiring continuity of the axial current and
are excluded from the bag interior so that chiral symmetry is realized in
the Wigner-Weyl mode inside the bag and in the Nambu-Goldstone mode outside.
As in Ref.~\cite{um2}, we use an
approximation of the chiral bag
model$^{\cite{myh}}$ where the mesonic cloud
surrounding the quark core is treated as a perturbation,
giving our model Lagrangian to lowest order in the meson fields as
\beqa
{\cal L}_{Bag} & = & \left( i\bar{q}(x)\!\! \not\!\partial q(x) - \frac{1}{4}
\sum_a F_{\mu\nu}^a(x) F^{\mu\nu a}(x) -B \right) \theta_V \nonumber \\
               &   & -\frac{1}{2}\bar{q}(x)\left( 1 + \frac{i}{f}
\vec{\lambda}\cdot\vec{\phi}(x)\gamma_5 \right) q(x)  \delta_S
+ \frac{1}{2} (\partial_{\mu} \vec{\phi}(x))^2\theta_{\bar{V}}
\label{eq:ONE}
\eeqa
where $q(x)$ and $\vec{\phi}(x)$ are the quark and octet meson
fields, respectively, and
\beq
\theta_V = \left\{ \begin{array}{c}
                  1,\: {\rm if\/}\, x \epsilon V \\
                  0,\: {\rm if\/}\, x\!\! \not\!\epsilon V
                  \end{array}
           \right. .
\label{eq:TWO}
\eeq
$\theta_{\bar{V}}$ is the compliment of $\theta_V$, $f$ is the appropriate
meson decay constant, (in this work we use $f_K = 1.09f_{\pi}$) and
$\delta_S = \delta(|\vec{x}\,|-R)$
couples the quarks to meson fields on the bag surface.
We will also give
masses to the quarks and thereby to the mesons
in our calculations.
We shall refer to the above Lagrangian when constructing effective meson
electromagnetic transition operators in Section~3.

The method we use to calculate the excited baryon masses in our approximation
of the
chiral bag has been described in detail in Refs.~\cite{um1} and \cite{mw}.
The extension of the model calculation to include the kaon cloud is
straightforward.
As discussed in Refs.~\cite{um1} and \cite{mw}, when calculating the effective
quark one-meson exchange diagrams, only those diagrams where the intermediate
baryon state, $|B'\ket$,
is the {\it same\/} as initial/final baryon state, $|B\ket$, are included
in the diagonalization
of the Hamiltonian and minimization of the energy (see Fig.~1 of
Ref.~\cite{um1}). Contributions to the baryon masses from those diagrams
where the intermediate baryon
state is {\it different\/} from initial/final baryon state are included
as corrections {\it after} the diagonalization and minimization procedure.
Consequently, when calculating the $\lams$ and
$\sigs$ masses {\it all\/} contributions from the kaon cloud are treated as
perturbative corrections to the masses and will therefore not affect the
hyperon wavefunctions obtained in Ref.~\cite{um2}.

Fig.~1 shows the spectrum of low-lying negative parity hyperon resonances
calculated in our model when only the pion field and
when both the pion and kaon fields are included in the meson cloud.
In the present calculation we consider contributions from all quark
one-gluon and effective one-pion
and one-kaon exchange interactions to the masses.
Here, as in Refs.~\cite{um2,um1,mw}, we do not correct for possible mass
differences between the initial/final and intermediate baryon states in the
meson cloud correction terms. In the following section and in Appendix~A we
will discuss this particular mass difference correction and present an
application to the lowest $\Lambda^*$ states.
The bag parameters used to fit the spectrum are
$B^{1/4}$ = 145 MeV, $Z_0$ = 0.25, $\alpha_s$ = 1.5 and $m_S$ = 250 MeV,
which are the same as in Ref.~\cite{um2}, where the meson cloud consists only
of pions, except that the zero-point/center-of-mass energy parameter,
$Z_0$, has been changed from 0.45 to 0.25.
This change in $Z_0$ is necessary because the kaon cloud lowers the masses of
all states by about 30 to 60 MeV relative to our results in Ref.~\cite{um2}
for a wide range of input parameters as exemplified in Fig.~1.
However, the wavefunctions of the
excited hyperon states shown in Fig.~1 remain the same as those obtained in
Ref.~\cite{um2}, since a change in $Z_0$ only affects the diagonal elements of
the bag Hamiltonian.
(The above change in $Z_0$ increases the stable bag radii by a few percent.)

It is instructive to compare our results with that of the NRQM calculation by
Isgur and Karl.$^{\cite{ik}}$ Table~1 is a summary of the masses of low-lying
negative parity hyperons predicted in the NRQM and in our model together with
the observed states.
With the exception of $\lamb$ to be discussed below, both models
can reproduce the established hyperon resonances reasonably well and give
similar predictions for states that have not been seen.
In particular, they are successful in describing the ordering of the
$J^P=\frac{5}{2}^-$
$\lams$ and $\sigs$ states correctly as well as the mass splitting between the
$J^P=\frac{3}{2}^-$ $\lama$ and $\Lambda(1690)$. In the $\sigs$ sector the
states
are not yet well determined experimentally but both models predict states near
1800 MeV and 1650 MeV with quantum numbers $J^P=\frac{3}{2}^-$ and
$\frac{1}{2}^-$, respectively.
However, from Table 1 alone it is clear that
the calculated hyperon states can not be unambiguously identified with
the observed resonances in the $I = 1$ sector.

Although both the NRQM and the chiral bag model predict similar
masses for the negative parity hyperon resonances, their wavefunctions
are very different as shown in Table 2, which lists the model
predictions for spin-flavor contents of the excited hyperons in percents.
(Note that in both models the $J^P=\frac{5}{2}^-$ $\lams$ and $\sigs$
are pure spin-quartet, flavor-octet states.)
Since the chiral bag model predicts an excited hyperon mass
spectrum very similar to the NRQM the
differences in the predictions for the spin-flavor contents of
these hyperon resonances between the two models become important and
should be explored experimentally as
discussed in Refs.~\cite{kms} and \cite{um2}.

\subsection{A modified meson cloud correction}

The most conspicuous state that both the NRQM and the above chiral bag
model calculation fail to reproduce is the $J^P =
\frac{1}{2}^- \lamb$ state, a possible $\bar{K}N$ bound state mentioned
in the introduction.
Recall that the NRQM predicts the lightest $J^P=\frac{1}{2}^-$ and
$\frac{3}{2}^-$ $\lams$ states to be mass-degenerate and
our model gives almost the same mass for these two states as well. In our
model,
the meson cloud contributions to baryon masses correspond to evaluating baryon
self-energy diagrams in which the initial baryon state couples to allowed
intermediate meson-baryon channels.
As mentioned above, it is necessary to correct these self-energy
diagrams for possible mass differences between the initial/final and
intermediate baryon states.
Therefore, we present here meson cloud corrections
which are modified by the mass
corrected dispersive effects and explore whether it can explain the observed
mass splitting between $\lama$ and $\lamb$.

The mass difference corrections to baryon self-energy diagrams have been
calculated in the chiral bag model within the perturbative meson cloud
approximation for
the ground state baryon masses$^{\cite{mbx}}$ using an effective Yukawa
model.$^{\cite{jaf2}}$ In our calculations, the
meson cloud contribution to the baryon masses, {\it i.e.}, the effective quark
one-meson exchange interaction matrix element, $\Sigma_B$, is written
as$^{\cite{mbx}}$
\beq
\Sigma_B = \sum_{i,j} \bra B|O(i)O(j)|B\ket
= \sum_{B'} \sum_{i,j} \bra B|O(i)|B'\ket \bra B'|O(j)|B\ket
\label{eq:THREE}.
\eeq
Here $i$ and $j$ are summed over the three quarks and the operator product
$O(i)O(j)$ is the
appropriate two-body (if $i\neq j$ )
one-meson exchange interaction operator which acts on
quarks $i$ and $j$ in the spin-flavor space.
Due to mass differences between
the states $|B\ket$ and $|B'\ket$, each term in the above sum over $B'$
should be multiplied by the real part of a correction factor
$\delta_L(BB')$, which, in the static approximation, is given by
\beq
\delta _{L}(BB') = \int_0^K \frac{dq\, q^{2(L+1)}}{\omega(\omega + m_{B'} -
m_{B} - i \epsilon)}
\left/ \int_0^K \frac{dq\, q^{2(L+1)}}{\omega^2}. \right. \label{eq:FOUR}
\eeq
Here $\omega=\sqrt{\mu^2+q^2}$ where $q$ is the meson momentum and $\mu$ is the
meson mass.
The momentum $K$ ($\sim$ 500 MeV) $^{\cite{mbx}}$ is
a cut-off reflecting the finite size of the source of the meson fields and is
determined by the normalization integral appearing in the denominator of
Eq.~(\ref{eq:FOUR}).
In the static approximation, $K$ is of the order of $1/R$ where $R$
is the bag radius. For the non-static case, the expression
$\omega + m_{B'} - m_B - i \epsilon$ in Eq.~(\ref{eq:FOUR})
should be replaced by $\omega + \sqrt{m^2_{B'} + q^2} - m_B - i \epsilon$,
and in our estimate
we use the same normalization as in the static case.
The relative intermediate meson-baryon angular momentum $L$ has the values of
$L=0$ for $B=\Lambda (\frac{1}{2}^-)$ and $L=2$ for $B=\Lambda (\frac{3}{2}^-)$
in our calculations.
In this effective Yukawa model, the imaginary part of $\delta_L(BB')$ gives the
hadronic decay width of the baryon $B$.

The sum over intermediate baryon states $|B'\ket$ in Eq.~(\ref{eq:THREE})
should, in principle, include all the observed baryon states in the allowed
meson-baryon channels for a given initial state $|B\ket$.
The ground state baryons are described by restricting the quarks to occupy only
the lowest $S$-state. In this case, almost all contributions to the baryon
self-energy diagrams come from octet and decuplet intermediate baryon ground
states because
higher excited $|B'\ket$ state contributions are negligible due to the large
mass differences $m_{B'}-m_B$. Therefore when $|B\ket$ are the ground state
baryons ({\it i.e.} when quarks occupy only the $S$-states), their
self-energies
may be evaluated by considering only the ground state intermediate states,
and in such a case
there are no ambiguities in assigning wavefunctions to each intermediate state
$|B'\ket$. Using this prescription Myhrer, Brown and Xu$^{\cite{mbx}}$ find
that the real part of multiplicative correction
factors, $Re\delta_L(BB')$,
are close to unity resulting in small
corrections to baryon self-energies and
help to improve the fit to the ground state baryon mass spectrum.

However, if the initial and final baryon state is an excited state,
it becomes necessary to include both ground state and excited baryon states
in $|B'\ket$, and this method runs into difficulties of a practical nature.
Even when we limit $|B'\ket$
to the quark space of either all three quarks in the lowest
$S$-state or two in the lowest $S$-state and
one in the lowest $P$-state ($P_{1/2}$ or $P_{3/2}$),
difficulties arise in the choice of the
calculated wavefunctions to be used in $|B'\ket$.
For example, consider contributions from the one-pion exchange
interaction to the mass of $\lams$. In this case possible intermediate baryon
states are the isospin one $\Sigma$ and their excited states,
and the $\sigs$ wavefunctions are needed to evaluate each term in the one-pion
exchange interaction matrix element, $\Sigma_B$.
However, as shown in Fig.~1, the $\sigs$ mass spectrum
is very poorly known experimentally$^{\cite{pdg}}$ and many states
predicted by quark models are not seen.
This makes it very difficult to assign calculated negative parity $\sigs$
wavefunctions to observed resonances and contributes to the theoretical
uncertainty in the resulting
$\lams$ mass. Another example of practical difficulties arises
in calculating the mass corrections due to the one-$K^+$ exchange.
This requires knowledge of our model wavefunctions for the
low-lying negative parity $\Xi^-$ states which we have not yet calculated.
Because of these difficulties, we chose to ignore corrections due to mass
differences between the initial/final and intermediate baryon states
when we calculate the baryon self-energy contributions to the
overall excited hyperon mass spectrum discussed in Section 2.1 and shown
in Fig.~1. Again, note that because the mass difference corrections apply only
to those one-meson exchange
diagrams where the intermediate baryon state is {\it different\/} from the
initial/final baryon state, they do not affect  our calculated wavefunctions
of excited hyperons which are used in Section 3.

We shall now make a very restricted application of the above mass corrections
to see if a better treatment of the
meson-baryon channels in the baryon self-energy diagrams might contribute to
an understanding of the mass splitting between $\lama$ and $\lamb$.
The lightest
$\Lambda(\frac{3}{2}^-)$ and $\Lambda(\frac{1}{2}^-)$ states
are identified with
$\lama$ and $\lamb$ and are denoted as $\Lambda(\frac{3}{2}^-)_3$ and
$\Lambda(\frac{1}{2}^-)_3$ in Table 1, respectively.
We shall estimate the magnitude of mass corrections in the spirit of the
work by Arima and Yazaki.$^{\cite{ay}}$
They investigated the problem of mass splitting between $\lama$
and $\lamb$ using a NRQM with mesonic degrees of freedom,
assuming that the two states have equal "bare" masses.
In this effective meson-baryon model the
hyperons acquire their physical masses through the
coupling to the meson-baryon channels.
They found that in order to generate a low mass for
the lightest $J^P=\frac{1}{2}^- \Lambda^*$-state
relative to the $\frac{3}{2}^-$ state, it was necessary to introduce
strong couplings to the intermediate $\bar{K}N$ and $\pi\Sigma$ channel.
This is also
similar in spirit to the work in the Cloudy Bag Model$^{\cite{cbag}}$
which unfortunately
was limited to $\lamb$ only. We consider it very important
to describe simultaneously both $\lama$ and $\lamb$ in any model of baryons
since the coupling in the $J^P = \frac{1}{2}^-$ and $\frac{3}{2}^-$ channels
are linked and both states couple to the $\pi\Sigma$ and $\bar{K} N$
meson-baryon channels.
When we consider the pure hyperon three quark states including one-gluon
exchange corrections, we find that the two lightest $\Lambda(\frac{3}{2}^-)$
and $\Lambda(\frac{1}{2}^-)$ states are almost mass degenerate$^{\cite{um2}}$,
the assumption made by Arima and Yazaki.$^{\cite{ay}}$
We find that the masses for these states remain nearly degenerate
when the pion cloud is included,$^{\cite{um2}}$
but the kaon cloud lifts this mass degeneracy as seen in Fig.~1
and discussed in Appendix~A. In our estimates  we shall use
$\Sigma_B$ of Eq.~(\ref{eq:THREE}) with the real part of the correction
factor of Eq.~(\ref{eq:FOUR}) and
consider only the case where $B=\lams$ and with
all the intermediate quarks in the $S$-state.
The ground state $N$ and $\Delta$ contributions
have different masses due to the one-gluon exchange interaction,
and to evaluate the largest corrections, we neglect the heavy
$\bar{K}\Delta$ and $\pi\Sigma^*(1385)$ intermediate states in our estimate
as in earlier calculations.$^{\cite{cbag,tor,ay,sg}}$
This means $B'$ includes only the octet baryons, {\it i.e.,}
$B'=N$ for the kaon cloud and $B'=\Sigma$ for the pion cloud.
Since we are calculating the meson cloud
contribution to the hyperon masses,
one should not use the physical $B$ and $B'$ masses,
but only their "bare" masses which
include the gluonic mass corrections$^{\cite{mbx}}$
that exist even in the chiral limit.
Additional details are given in Appendix~A.

Our model estimate confirms the findings of Arima and Yazaki$^{\cite{ay}}$
that quark models with a $qqq$ structure for hyperons have
difficulties in explaining the $\lama$ and $\lamb$ mass difference.
Further examinations regarding the nature of
$\Lambda(1405)$ are necessary, and we stress again that it is imperative to
treat both
$\Lambda(1405)$ and $\Lambda(1520)$ in the same model and then repeat,
for example, Veit {\it et.al.}'s Cloudy Bag Model calculation$^{\cite{cbag}}$
starting with the "bare" masses of $\Lambda(1405)$ and $\Lambda(1520)$
about equal and use the coupling to the meson-baryon channels to
get the observed hyperon masses.
We note that the Cloudy Bag Model allows for a quark-meson four point
interaction in the bag volume which
generates an S-wave meson-baryon contact interaction.
In their model this specific interaction, which is of second
order in $f_{\pi}^{-2}$
and is {\it not} included in our model, is responsible for the
remarkable lowering of
the $\lamb$ mass. It would be interesting to investigate if this
second order quark-meson interaction
will affect the mass of the $\lama$ equally strongly. Such an investigation
might give us some further clues to the question of whether the quark content
of $\Lambda(1405)$ differs from that of $\Lambda(1520)$ or
not.\footnote{In {\it all} of the Cloudy Bag Model
calculations, the $\lamb$ is {\it assumed} to be a flavor singlet state.}
We find that although the "bare masses" ({\it i.e.} masses before meson cloud
corrections) of our two {\it lightest} $J^P = \frac{1}{2}^-$ and
$\frac{3}{2}^-$ hyperon states are about equal ($\sim 1695$ MeV with the bag
parameters used in Fig.~1), their flavor contents are different with
$\Lambda(\frac{1}{2}^-)_3$
being almost equal mixtures of flavor singlet and flavor octet states,
(see Table 2).

\subsection{Strong decay widths and coupling constants}

In this subsection we estimate the hadronic decay widths of the two lightest
excited hyperons, given by the imaginary part of Eq.~(\ref{eq:FOUR}), and the
coupling constants of $\lamb$ into the meson-baryon channels. For the hadronic
decay widths we use the observed physical mass differences and our
"non-static" expression
in Eq.~(\ref{eq:FOUR}) to get the correct centrifugal factor, using
a bag radius $R$ of 1.2 fm and a strange quark mass $m_S$ of 250 MeV.
Also, as discussed in Appendix~A, we do not restrict the sum over the quark
indicies to the case $i \neq j$ in Eq.~(\ref{eq:THREE}).
Then the hadronic decay width of $\lama$ into the $\pi\Sigma$ channel is
given by
\begin{equation}
\Gamma_{\pi\Sigma}[\lama] = 5 \left( \frac{a}{\sqrt{2}}
- \frac{b}{2 \sqrt{5}} - \frac{c}{2}  \right)^2 \times 7.7\: {\rm MeV}.
\label{eq:FIVE}
\end{equation}
Here the strength of the quadrupole transition operator
$k^{\pi}_{SAAS}$ of Ref.~\cite{um1}, Table II, is used to find the width.
A similar quadrupole transition operator for the kaon cloud,
$k^K_{SA'A'S}$, ($A'$ indicates a massive $s$-quark in the $P_{3/2}$ state),
is used to determine the hadronic width of $\lama$ into the $\bar{K}N$ channel
which we find to be
\begin{equation}
\Gamma_{\bar{K}N}[\lama]  = \frac{5}{2} \left(a + \sqrt{2}c \right)^2
\times 2.4 \: {\rm MeV}.
\label{eq:SIX}
\end{equation}
For $\Lambda(1405)$ the hadronic width is given by
\begin{equation}
\Gamma_{\pi\Sigma}[\lamb] = \left( a' + \frac{b'}{\sqrt{2}} -
\frac{c'}{\sqrt{2}} \right)^2 \times 19.3\: {\rm MeV},
\label{eq:SEVEN}
\end{equation}
where the magnitude of the width is determined by the
monopole transition operator for the pion cloud, $ m^{\pi}_{SPPS}$,
of Ref.~\cite{um1}.
Here $a, b, c$ and $a', b', c'$ are the spin-flavor coefficients for the
excited hyperon wavefunctions in
the $j-j$ coupled basis. For $\lama$ they are defined as
\beq
|\lama\ket \equiv a|{\bf 1};SSA\ket' + b|{\bf 8};SSA\ket + c|{\bf 8};SSA\ket'
+ d|{\bf 8};SSP\ket
\label{eq:EIGHT}
\eeq
where $S\equiv S_{1/2}, P\equiv P_{1/2}$ and $A\equiv P_{3/2}$,
and the corresponding coefficients for $\lamb$ are
\beq
|\lamb\ket \equiv a'|{\bf 1};SSP\ket' + b'|{\bf 8};SSP\ket +
c'|{\bf 8};SSP\ket' + d'|{\bf 8};SSA\ket.
\label{eq:NINE}
\eeq

Using the calculated wavefunctions for $\lama$ and $\lamb$ given in Eqs.~(5)
and (6) of Ref.~\cite{um2}
\footnote{Eq.~(6) in Ref.~\cite{um2} contains a misprint: 0.03 should be
0.003.}
together with its appendix, one finds the following values for these
coefficients;
\beq
a=+0.95\hspace{0.5cm} b=-0.21\hspace{0.5cm} c=+0.21\hspace{0.5cm}
d=0.07
\label{eq:TEN}
\eeq
and
\beq
a'=+0.77\hspace{0.5cm} b'=+0.15\hspace{0.5cm} c'=+0.45\hspace{0.5cm}
d'=-0.43
\label{eq:ELEVEN}
\eeq
With these values for the spin-flavor coefficients we find the total
$\Lambda(1520)$ decay width to be 23.6 MeV
compared to the measured width of 15.6 MeV.$^{\cite{pdg}}$
Note that if $\Lambda(1520)$ is a pure flavor singlet state ($a=1, b=0, c=0$
and $d=0$) the relative branching ratio
$\Gamma_{\pi\Sigma}[\lama] / \Gamma_{\bar{K}N}[\lama]$
increases from 14.4/9.2 $\approx$ 1.6 to approximately 7.7/2.4 $\approx$ 3.2.
For $\lamb$ we find $\Gamma_{\pi\Sigma}[\lamb] \approx 6 $ MeV,
whereas if $\lamb$ is a pure flavor singlet then
Eq.~(\ref{eq:SEVEN}) gives a width of about 19.3 MeV.
We note that the values of
these widths depend somewhat on the value of the bag radius $R$.
For example, if $R = 1.175$ fm then the total widths for the $\lama$ decreases
to 19 MeV, a value closer to the measured width, but
$\Gamma_{\pi\Sigma}[\lamb]$
changes only by 1 MeV to 5 MeV. The value of the $\lamb$ width is difficult
to extract from experimental data since this hyperon,
a resonance only 27 MeV below the $K^- p$ threshold,
can only be  seen in the $\pi \Sigma$ channel. However,
our calculated hadronic width for $\lamb$ are clearly too small, a result which
will be reflected in a too small hadronic coupling constant as discussed in the
following paragraph.
The latest analysis of the
experimental $\pi^-\Sigma^+$ mass spectrum by Dalitz and Deloff$^{\cite{dd}}$
gives a value of 50$\pm$2 MeV for this width.

The magnitude of the coupling constants for the hyperon $Y^*$ coupling to the
$\bar{K} N$ and $\pi\Sigma$ channels can also be calculated in the chiral bag
model. For $\Lambda(1405)$ we find the ratio of these coupling constants to be
\begin{equation}
\frac{G_{\lamb \bar{K} N}^2}{G_{\lamb \pi \Sigma}^2} = \frac{2}{3}
\frac{m^K_{SP'P'S}}{m^{\pi}_{SPPS}} \frac{(a' + \sqrt{2}c')^2}
{(a' +  b'/\sqrt{2} - c'/\sqrt{2})^2}.
\label{eq:TWELVE}
\end{equation}
where $m^K_{SP'P'S} / m^{\pi}_{SPPS}$ is the ratio of the strength of the
monopole transition operator for the kaon cloud to that of the pion cloud,
which depends on $R$ and $m_S$. (Similar to
above, $P'$ denotes a massive $s$-quark
in the $P_{1/2}$ state.) In Eq.~(\ref{eq:TWELVE}) we use the same convention as
in Table~6.6 of the compilation by Dumbrajs {\it et.al.}$^{\cite{dum}}$
where $G^2_{\lamb\pi\Sigma}$ is defined in their
Eq.~(6.4).\footnote{The definitions of strong
coupling constants used in Ref.~\cite{dum} are not always consistent.
Also the footnote (a) to Table~6.6 in this reference is incorrect.
The ratio of coupling constants should {\it increase} if $g^2$ were used as
correctly stated in footnote (a) of Table~6.14 of Nagels
{\it et.al.}$^{\cite{nagels}}$}
Note that from
Eqs.~(\ref{eq:FIVE}), (\ref{eq:SIX}), (\ref{eq:SEVEN}) and (\ref{eq:TWELVE})
one sees that neither the $|{\bf 8};SSP\ket$ component of $\lama$ nor
the $|{\bf 8};SSA\ket$ component of $\lamb$ couple to the $\bar{K}N$ or
the $\pi\Sigma$ channels.
 From Eq.~(\ref{eq:TWELVE}) it is clear that the ratio of the
two $\lamb$ coupling
constants equals 2/3 only if: {\it (i)} $\Lambda(1405)$ is a flavor singlet
($a' = 1$; $b' = c' = d' = 0$)
and {\it (ii)} we are in the $SU(3)$ limit where the pion and kaon masses
as well as the $u$-, $d$- and $s$-quark masses
are equal, so that the ratio $m^K_{SP'P'S}/m^{\pi}_{SPPS}=1$.
Otherwise, the ratio of the coupling constants will be
different from 2/3.
The flavor octet part of $\Lambda(1405)$
will increase the value of the ratio from 2/3,
whereas the meson cloud corrections to this "bare" quark model value tend
to make the ratio much smaller than 2/3 due to
the very different kaon and pion masses.
With our values for the spin-flavor coefficients $a'$, $b'$ and $c'$ given in
Eq.~(\ref{eq:ELEVEN}) we find the ratio of
coupling constants in Eq.~(\ref{eq:TWELVE}) to be $\approx$ 4.3 in the $SU(3)$
limit. Including the physical meson masses,
which corresponds to $m^K_{SP'P'S}/m^{\pi}_{SPPS} \approx 0.11$ at $R$ = 1.2 fm
and $m_S$ = 250 MeV, we find
this ratio reduced to approximately 0.47.
It would be very useful to reanalyze the $\bar{K} N$ scattering data to see
if this ratio of coupling constants for $\lamb$ is much smaller or larger
than 2/3. If one uses both Table~6.6 in Dumbrajs {\it et.al.}$^{\cite{dum}}$
and Table~6.14 of an earlier data analysis compilation by Nagels
{\it et.al.},$^{\cite{nagels}}$ it is seen that even a small value for the
ratio of the coupling constants is not ruled out.
However, in our model the quark-meson coupling is at the bag surface,
and the meson fields have a sharp cut-off, which is unphysical.
Our meson cloud contribution to the ratio $m^K_{SP'P'S}/m^{\pi}_{SPPS} \approx
0.11$ might therefore be unrealistic.

Using Eq.~(6.4) of Dumbrajs {\it et.al.}$^{\cite{dum}}$ and the value we found
above for the hadronic width of $\lamb$, the magnitude of the coupling constant
$|G_{\lamb\pi\Sigma}|$ is found to be 0.698 with $R = 1.2$ fm and $m_S = 250$
MeV. Here $G_{\lamb\pi\Sigma}$ is the charge independent coupling to
the $\pi\Sigma$. The ratio $G^2_{\lamb KN}/G^2_{\lamb\pi\Sigma} =
0.43$ then determines the coupling of $\lamb$ to the $\bar{K}N$ channel,
$|G_{\lamb KN}| = 0.46.$ If $\lamb$ were a pure flavor singlet state,
the magnitude
of the $\lamb\pi\Sigma$ coupling constant increases to $|G_{\lamb\pi\Sigma}|_1
= 1.211$ while the corresponding coupling to the $\bar{K}N$
channel decreases to $|G_{\lamb KN}|_1 = 0.327$.

In addition to these coupling
constants for $\lamb$, we have also calculated the magnitude of the strong
coupling constants $G_{\lamz \bar{K} N}$ and $G_{\sigz \bar{K} N}$ within our
model as shown in Appendix~B.
These hyperon coupling constants are defined for each charge state to conform
to the notation of Table~6.3 and Section~2.3 of Dumbrajs
{\it et.al.}$^{\cite{dum}}$. With $R = 1.125$ fm for the $\lams$ and $\sigs$
and $m_S = 250$ MeV, we find that the chiral bag model gives
$|G_{\lamz \bar{K} N}| = 9.68$ and $|G_{\sigz \bar{K} N}| = 3.23$, values
which are somewhat smaller than those quoted
in Table~6.3 of Dumbrajs {\it et.al.}$^{\cite{dum}}$ However, these values for
$\lamz \bar{K}N$ and $\sigz \bar{K}N$ coupling constants are well within the
range reported in a more recent compilation by Adelseck and
Saghai.$^{\cite{as}}$ We note that a recent calculation in the bound state
approach to the Skyrme model by Gobbi {\it et.al.}
gives $G_{\lamz \bar{K} N} = -9.93$ and $G_{\sigz \bar{K} N} = +3.43$ when
evaluated with a pion mass term in the model lagrangian.$^{\cite{gobbi}}$

\section{Hyperon Radiative Decay Widths: Meson Cloud Contribution}

As stated above, it is not
possible to distinguish the NRQM from the chiral bag model using
the  results presented in Tables~1 and 2. Instead, one needs to compare
calculated observables which are sensitive to the predicted hyperon
wavefunctions of Table~2. One such observable
is the radiative decay widths of excited hyperons decaying into their ground
states. In this section we use our simple two phase model of baryons,
described by the
Lagrangian in Eq.~(\ref{eq:ONE}), to extract meson transition electromagnetic
currents and evaluate meson cloud corrections to the total hyperon radiative
decay widths.

Before presenting the details of the calculation it should be emphasized that
in our model the quark-meson coupling at the bag surface, determined by the
requirement of a continuous axial current in the chiral symmetry limit, is
{\it pseudoscalar\/} implying that the
photon only couples to the mesons through the kinetic energy term.
Therefore, contributions to the meson electromagnetic current comes only from
those
diagrams where the photon couples to the mesons in flight as shown in Fig.~2c.
The Cloudy Bag Model,$^{\cite{cbag}}$ uses, in addition to the four point
quark-meson interaction, a derivative quark-meson coupling in the bag volume.
This means that the photon can couple to the mesons inside the quark core
and allows for a quark-meson-photon contact interaction (Figs.~2a and 2b).
We ignore contributions to the decay widths from those diagrams
shown in Fig.~2d, where the intermediate baryon radiates a photon while
the meson is in flight. In principle, this diagram should be included but its
effect is small compared with the process shown in Fig.~2c, roughly by the
ratio of meson to baryon masses. As we shall show, the contribution from
the diagram Fig.~2c to the total hyperon radiative width is small and therefore
ignoring Fig.~2d will not have any numerical consequence in this calculation.
This approximation is also invoked by
Zhong {\it et.al.}$^{\cite{cloudy}}$ who used a chiral $SU(3)$ version of the
Cloudy Bag Model to calculate the branching ratios of the $K^-p$ atom.
In their calculation, the $\lamb$, which they assume to be a purely flavor
singlet state, gives a relatively small contribution to the
radiative decay width of the $K^-p$ atom. The reason is that
$\bar{K}N$ atomic system is close to the $\bar{K}N$ threshold which
is at the upper tail of the $\lamb$ width,
and the $\lamb$ coupling strength is therefore
considerably reduced in the atomic branching ratios.

Having defined the model Lagrangian, meson electromagnetic transition currents
can be extracted in the usual manner by introducing the minimal coupling
prescription $\partial_{\mu} \rightarrow \partial_{\mu}\pm ieA_{\mu}$.
The total decay width receives contributions both from the
quark core and the meson cloud and, using the notation of Ref.~\cite{kms},
is given by
\beq
\Gamma_{J_i J_f} = \frac{2k}{2J_i + 1} \sum_{m_i m_f} \,
\sum_{\lambda=\pm 1} \left| \bra J_f m_f \left|
\hat{\epsilon}^*_\lambda(\hat{k})
\cdot (\vec{I}_q + \vec{I}_m) \right| J_i m_i \ket \right|^2.
\label{eq:THIRTEEN}
\eeq
Here $\hat{\epsilon}$ is the photon polarization vector and $\vec{I}_q$ and
$\vec{I}_m$ are defined as
\beqa
\vec{I}_q & \equiv & \int_{0}^{R}\! d^3r\, \vec{J}_q(\vec{r})
e^{-i\vec{k}\cdot\vec{r}}
\label{eq:FOURTEEN}
\\
\vec{I}_m & \equiv & \int_{R}^{\infty}\! d^3r\, \vec{J}_m(\vec{r})
e^{-i\vec{k}\cdot\vec{r}},
\label{eq:FIFTEEN}
\eeqa
where $R$ is the bag radius.
In Eqs.~(\ref{eq:FOURTEEN}) and (\ref{eq:FIFTEEN}), $\vec{J}_q$ and $\vec{J}_m$
are the quark and meson electromagnetic current operators, respectively, and,
as in Refs.~\cite{kms,um2},
we take the photon momentum $k$ to be given by
the observed mass difference between the initial and final hyperon states.
The current $\vec{J}_m$ is generally a two body operator such that
the transition operator given in Eq.~(\ref{eq:FIFTEEN}) has the structure
$\vec{I}_m = \sum_{i,j} O(i)O(j)$ which is similar to the effective quark
one-meson exchange interaction operator in Eq.~(\ref{eq:THREE}).
Thus, for example, the meson cloud contribution to
the matrix element for the radiative decay of $\lamb$ into $\Sigma^0$
is written as
\beq
\bra\sigz|\vec{I}_m|\lamb\ket = \sum_{B'}\sum_{i,j}
\bra\sigz|O(i)|B'\ket\bra B'|O(j)|\lamb\ket.
\label{eq:SIXTEEN}
\eeq
As a result, we are faced with the problem of the
choice of wavefunctions for the intermediate baryon states, a problem similar
to the one already addressed in the previous section.
In this work we use
only the ground state octet and decuplet baryons with
all three quarks in the $S$-state for intermediate $|B'\ket$ states
and, as a first approximation, neglect
baryon mass differences as well as recoil corrections.
This means the possible
intermediate baryon states included in Eq.~(\ref{eq:SIXTEEN}) are
$B' = p, \Xi^-,
\Sigma^+, \Sigma^-,\Delta^+, \Xi^{-*}, \Sigma^{+*}$ and $\Sigma^{-*}$.
In Appendix~C we present some explicit examples
of the meson electromagnetic currents,
$\vec{J}_m$, and give expressions for the corresponding transition
operators $\vec{I}_m$.

We find that contributions from
$\pi^+\Sigma^- $ and $\pi^-\Sigma^+$ as well as $\pi^+\Sigma^{-*}$ and
$\pi^-\Sigma^{+*}$ intermediate states cancel each other when evaluating the
the matrix element $\bra\lamz|\vec{I}_m|\lams\ket$ for the radiative decay
of $\lams$ into $\lamz\gamma$.
The reason is that the strong coupling constants
for the processes $\pi^-\Sigma^+ \rightarrow \Lambda$ and
$\pi^+\Sigma^- \rightarrow \Lambda$ (as well as $\pi^{\pm}\Sigma^{\mp}
\rightarrow \Lambda^*$) have the {\it same} sign
whereas the $\pi^+$ and $\pi^-$
transition electromagnetic currents connecting the states $\lams$ and $\Lambda$
have the {\it opposite} sign thus cancelling the contributions from
$\pi^+\Sigma^-$ and $\pi^-\Sigma^+$ intermediate states in the matrix element
for the radiative widths.
Consequently, in this model, the pions are effectively spectators
for $\lams \rightarrow \lamz \gamma$ decays in
contrast to $\lams \rightarrow \sigz \gamma$ decays where both the
kaon and pion clouds contribute to the total decay widths.
Furthermore, there are no contributions from the $K^-\Delta^+$ intermediate
state to the radiative decays of $\lama$ and $\lamb$ for the following reason.
If the initial excited hyperon is to
radiate through the $K^-$ cloud, a strange quark must initially
be in an excited state
implying that both the $u$- and $d$- quarks are
in $S$-states and form an antisymmetric spin-flavor state.
Therefore, the initial hyperon state cannot couple to the
$\Delta^+$ intermediate state with a totally symmetric spin-flavor
wavefunction.

The results of our calculations
are summarized in Table~3 where we present
the separate incoherent
quark core and meson cloud contributions as well as
the total hyperon radiative decay widths.
As in Ref.~\cite{um2}, we use $m_S =
250$ MeV and $R = 1.125$ fm to calculate these widths.
It is clear that the meson cloud corrections to the
decay widths are negligible except in the case of the decay
$\lamb \rightarrow \Sigma^0 + \gamma$ where the
width decreases from 2.22 keV to 1.85 keV
when the meson cloud corrections are included.
The meson cloud contributions to the widths for the decay
process $\Lambda^* \rightarrow \Lambda + \gamma$ is much smaller than those for
$\Lambda^* \rightarrow \Sigma^0 + \gamma$ decays
as expected due to the heavy kaon mass, {\it i.e.},
the meson cloud contributions to the radiative decay widths, though small, come
from the pion cloud.

These results are a direct consequence of the weak meson field approximation to
the chiral bag model. In this approximation,
the meson fields are localized just outside the bag surface and
their strength decreases very rapidly as one moves away from
the quark source on the surface.
Numerically, the smallness of the meson cloud contributions
come from the small values of integrals $I_1$ to $I_6$
defined in Eqs.~(C-17) to (C-22) in Appendix~C.
These are small due to
the damping factor of $e^{-2\mu Rx}$ appearing in the integrands,
which means
only the region near the (sharp) bag surface contribute to the
integrals. In fact, this can be seen in
the radial charge density of the $\Lambda$, calculated
in the Cloudy Bag Model  by Kunz, Mulders and Miller.$^{\cite{kmm}}$
They find a vanishing charge density at a distance of about 0.5 fm
outside the bag cavity  (see Fig.~2 of Ref.~\cite{kmm}).
This situation is similar to the description of excited $\Lambda$ in our model
although the two models will
probably disagree in their predictions of the radiative decay widths due to
different meson-photon couplings.

\section{Discussion and Conclusion}

In this work we have calculated the masses of low-lying negative parity
hyperons in a perturbative approximation to the
chiral bag model incorporating both pion and kaon fields
in the meson cloud and found that the resulting mass spectrum is very similar
to
the one predicted by the NRQM calculation of Isgur and Karl.$^{\cite{ik}}$
Not only do both models reproduce this mass spectrum to some extent, but
they also predict yet unobserved states sharing the same quantum numbers.
However, both models fail to reproduce the observed
mass splitting between $\lama$ and $\lamb$.
Without the kaon cloud, our model, like the NRQM,
finds the lightest $J^P=\frac{3}{2}^-$ and $\frac{1}{2}^-$ $\lams$ states to be
degenerate in mass. A more careful treatment
of the baryon self-energy diagrams takes the mass difference between
the initial/final and the intermediate baryons into consideration.
An estimate
of this mass difference correction indicates that it is the large
flavor-singlet component in the $\lama$ which lowers its mass below
that of $\lamb$ when the kaon cloud is included in our model.

We have also calculated the hadronic widths of $\lama$ and $\lamb$ and found
that
the total width of $\lama$ agrees well with experiment whereas the prediction
for the $\lamb$ width is too small relative to the currently accepted value.
If we include the effects of the kaon and pion clouds, the
ratio of the square of coupling constants of $\lamb$
coupling to the $\bar{K}N$ and $\pi\Sigma$ channels was also found to be
small compared to most
data analysis.$^{\cite{dum,nagels}}$
We find in this paper that the chiral bag model gives values for
the $\lamz KN$ and the
$\sigz KN$ coupling constants consistent with the latest compilation
by Adelseck and Saghai$^{\cite{as}}$ and the Skyrme model calculation by Gobbi
{\it et.al.}$^{\cite{gobbi}}$
In view of this it seems like our dominant $qqq$ assumption regarding
$\lamb$ has some difficulties. However,
we note that the only
reliable quantity in an analysis of $\lamb$ is the strength of the residue at
the $\lamb$ pole which can only be reached via a dispersion calculation
analysis of the experimental data.
Current $\bar{K}N$ and $\pi\Sigma$
scattering data are presently
too crude to allow for a reliable dispersion analysis.

In addition to the hadronic widths of the lightest $J^P = \frac{3}{2}^-$ and
$\frac{1}{2}^-$ hyperons, we evaluated the meson cloud
corrections to the radiative decay widths of $\lama$ and $\lamb$ which we claim
to be a very good observable to test various models of hadrons.
In our model the radiative decay of an
excited hyperon proceeds mainly through the radiation of a photon by
the excited valence quarks
and almost no radiation occurs from the meson cloud.
To examine this further, one should calculate the
excited baryon mass spectra with the
non-perturbative (topological) chiral bag model,$^{\cite{lbag}}$ which is
technically very challenging.
The model we use in this work is the weak meson
field approximation to the topological bag model,
which enhances the role of the
quarks in the bag cavity while ignoring the non-perturbative effects of the
meson cloud. Another extreme approximation to the topological bag model
is the soliton description of baryons without any explicit quark degrees
of freedom.
The Skyrme model is a prototype of such soliton models of baryons, and
hyperon resonances have been studied using this
model.$^{\cite{skyrme}}$
A calculation of the radiative
decay widths of excited hyperons in the Skyrme model for comparison with
predictions from models of baryons containing explicit quark degrees of
freedom
is currently in progress by one of the authors.$^{\cite{yu}}$

In Ref.~\cite{bl}, Burkhardt and Lowe extract the radiative decay widths of
$\lamb$ decaying into ground states $\sigz$ and $\lamz$ using the measured
branching ratios for the radiative decay of the $K^-p$ atom.$^{\cite{white}}$
They use a pole model to calculate the $K^-p$ atom radiative decay branching
ratios in order to determine the $\lamb$ radiative decay widths.
The $\lamb$ coupling constant to the $\bar{K}N$ channel and the
$\lamz \bar{K}N$ and $\sigz \bar{K}N$ coupling constants
are among the input parameters of this analysis.
We find a small
value of the $\lamb \bar{K}N$ coupling constant, a reflection of the small
$\lamb$ hadronic widths obtained through Eq.~(\ref{eq:SEVEN}). A pole model
analysis with this small coupling gives
$\lamb$ radiative decay widths
larger than predictions reported in the literature.$^{\cite{jl}}$
This together with our unsuccessful attempt in
obtaining a $\lama$ and $\lamb$ mass splitting and with the
Cloudy Bag Model results of going to the next order in the meson
field coupling, indicate that the $\Lambda(1405)$ structure is more than
a simple three quarks state.
In fact, it has been known for some time that the real part of the $K^-p$
scattering length extracted from the $1S$ level shift of the kaonic hydrogen
atom and from $\bar{K}N$ scattering have opposite signs. A recent examination
of this "kaonic hydrogen puzzle" by Tanaka and Suzuki$^{\cite{ts}}$ using two
different models seem to favor the one assuming a two-body composite system for
the $\lamb$.

In summary, with the exception of $\lamb$, the chiral bag model describes
the mass spectrum of the negative parity hyperons reasonably well
which, together with the earlier success of describing the $N^*$ and $\Delta^*$
negative parity mass spectrum,$^{\cite{mw,um1}}$
makes this model a strong competitor to the NRQM.
We have also calculated the partial and total
hadronic widths of $\lama$ and find them to be close to the
experimentally observed ones.
Furthermore, we find that the electromagnetic decay widths of the two lightest
excited hyperons
in this model are much smaller than the ones calculated in the
NRQM.$^{\cite{dhk}}$
We stress that to understand the difference between the structures of $\lama$
and $\lamb$ hyperons and the nature of their observed $\vec{L}\cdot\vec{S}$
splitting, both states, which are close to the $\bar{K}N$ threshold, should be
examined within the same model.
\vskip 1in
\noindent {\large\bf Acknowledgements}
\vskip 0.2in
We thank G.E.~Brown for encouraging and inspiring this
collaboration and Jim Lowe for numerous correspondences.
One of us (YU) would like to thank the University of South Carolina at Columbia
for kind hospitality.
\vfill\eject
%
%
\appendix
\setcounter{equation}{0}
\renewcommand{\theequation}{\thesection-\arabic{equation}}
\section{Hyperon Mass Corrections}

Here we discuss details regarding
the estimates of the mass correction factor ({\it i.e.} the real part of
$\delta_L(BB')$) in Eq.~(\ref{eq:FOUR}), which multiply
the spin-flavor matrix elements of Eq.~(\ref{eq:THREE}).
We apply this mass correction to $\Lambda(\frac{3}{2}^-)_3$ and
$\Lambda(\frac{1}{2}^-)_3$ and discuss why our model gives the unexpected
mass ordering of these two lightest excited hyperons as shown in Fig.~1.

In Eq.~(\ref{eq:FOUR}) we use the "bare" mass for the initial/final baryon
state, $m_B$, which for
$B=\Lambda(\frac{3}{2}^-)_3$ and $\Lambda(\frac{1}{2}^-)_3$ are about equal and
are approximately 1695 MeV.
This "bare" mass value is not too different from what is needed in the
calculations of Veit {\it et al.}$^{\cite{cbag}}$,
Siegel and Weise$^{\cite{sw}}$ and
Arima and Yazaki$^{\cite{ay}}$ and is a result
of our bag model calculation including all one-gluon exchange interaction
terms, but excluding all meson cloud contributions.$^{\cite{um2}}$ The "bare"
ground state baryon octet masses taken from Ref.~\cite{mbx} are used for the
intermediate baryon masses, $m_{B'}$.
In general, $m_{B'} - m_B =  - \Delta < 0$, which gives our "static"
estimate for the mass corrections. When $B=\Lambda(\frac{1}{2}^-)$,
Eq.~(\ref{eq:FOUR}) gives
$Re\delta_0(\Lambda(\frac{1}{2}^-)_3N) \approx 9$ for the
$\bar{K}N$ $L=0$ intermediate state and the correction factor for the
$\pi \Sigma$ intermediate state is
$Re\delta _0(\Lambda(\frac{1}{2}^-)_3\Sigma) \approx 3$.
For $B=\Lambda(\frac{3}{2}^-)$, we use the "non-static" estimate since the
baryon recoil will be particularly important for the $\bar{K}N$ intermediate
state with $L=2$. In this case,
the corresponding correction factors are found
to be $Re\delta _2(\Lambda(\frac{3}{2}^-)_3N) \approx 5$
and $Re\delta_2(\Lambda(\frac{3}{2}^-)_3\Sigma) \approx 4$ for the $L=2$
meson-baryon intermediate states.
The value of the correction factor $Re\delta_2(\Lambda(\frac{3}{2}^-)_3N)$
is very sensitive to the "bare" baryon mass difference $m_{B'} - m_{B}$
due to the closeness of the $\bar{K}N$ threshold and the
$L=2$ centrifugal factor.

Using these correction factors in Eq.~(\ref{eq:THREE}), we find
$\Lambda (\frac{3}{2}^-)_3$ to be lower in mass than
$\Lambda (\frac{1}{2}^-)_3$ contradicting observations if
we were to identify $\lama$ with $\Lambda(\frac{3}{2}^-)_3$ and $\lamb$ with
$\Lambda (\frac{1}{2}^-)_3$.
The reasons are inherent in our model and are as follows.
For $\Lambda(\frac{3}{2}^-)_3$, the $\bar{K} N$ and the $\pi \Sigma$
intermediate state couplings in Eq.~(\ref{eq:THREE}) operates only through
the $k^K_{SA'A'S}$ and $k^{\pi}_{SAAS}$ quadrupole transition operators,
(see Table II of Ref.~\cite{um1}). Here, as in Section~2.2, $A'$ denotes a
massive $s$-quark in the $P_{3/2}$ state.
In our model calculations, the wave function for the lightest
$J^P = \frac{3}{2}^-$ $\lams$ state was found to be$^{\cite{um2}}$
\beqa
|\Lambda(3/2^-)_3\ket & = & -0.95\,|^2 1_{3/2}\ket - 0.09\,
|^4 8_{3/2}\ket + 0.29\,|^2 8_{3/2}\ket
\label{eq:BONE}
\eeqa
which is clearly dominated by the flavor singlet component.
The pionic cloud contribution to the $J^P=\frac{3}{2}^-$ states,
$H_{\pi}$, is given by a 3$\times$3 matrix defined by
\beqa
H_{\pi}
& \equiv &
\left( \begin{array}{ccc}
\bra ^2 1_{3/2}|O_{\pi}|^2 1_{3/2} \ket &
\bra ^2 1_{3/2}|O_{\pi}| ^4 8_{3/2} \ket &
\bra ^2 1_{3/2}|O_{\pi}| ^2 8_{3/2} \ket \\

\bra ^4 8_{3/2}|O_{\pi}|^2 1_{3/2} \ket &
\bra ^4 8_{3/2}|O_{\pi}|^4 8_{3/2} \ket &
\bra ^4 8_{3/2}|O_{\pi}|^2 8_{3/2} \ket \\

\bra ^2 8_{3/2}|O_{\pi}|^2 1_{3/2} \ket &
\bra ^2 8_{3/2}|O_{\pi}|^4 8_{3/2} \ket &
\bra ^2 8_{3/2}|O_{\pi}|^2 8_{3/2} \ket
       \end{array}
\right)
\label{eq:BTWO}
\eeqa
where $O_{\pi}$ is the effective quark one-pion exchange interaction operator.
The matrix in Eq.~(\ref{eq:BTWO}) is similar to Eqs.~(2) in Ref.~\cite{um2}.
With a bag radius of 1.2 fm and strange quark mass of 250 MeV the matrix
elements in $H_{\pi}$ are (the numbers are in MeV)
\beqa
H_{\pi} & = &
\left( \begin{array}{ccc}
      -104&  -7&   22\\
       -7 & -41&   21\\
        22&  21&  -79
       \end{array}
\right)
\label{eq:BTHREE}
\eeqa
and the corresponding kaon cloud contribution is
\beqa
H_{K} & = &
\left( \begin{array}{ccc}
       -34&    1&    2\\
         1&  -25&    7\\
         2&    7&  -30\\
       \end{array}
\right)
\label{eq:BFOUR}
\eeqa
The intermediate baryon octet ground states (with all three quarks in
the $S$ state) contribute to the matrix elements in Eqs.~(\ref{eq:BTHREE}) and
(\ref{eq:BFOUR}).
The meson cloud contributions to the octet components of
$\Lambda(\frac{3}{2}^-)_3$ and $\Lambda(\frac{1}{2}^-)_3$ states keep
these two lowest $J^P=\frac{3}{2}^-$ and $\frac{1}{2}^-$ states approximately
mass degenerate.
Therefore, in this discussion we shall concentrate on the contribution from
the pure flavor singlet component which is responsible for
the unexpected  mass ordering of the lowest $J^P=\frac{3}{2}^-$ and
$\frac{1}{2}^-$ states.

The pure flavor singlet mass contributions from the pion cloud, denoted as
$\bra ^2 1_{3/2}|O_{\pi}|^2 1_{3/2} \ket$ in
Eq.~(\ref{eq:BTHREE}), is $-104$ MeV.
This contains a contribution from the $\pi\Sigma$ intermediate state of
$5Re\delta_2(\Lambda(\frac{3}{2}^-)_3\Sigma)k^{\pi}_{SAAS} = -16.6$ MeV
when $Re\delta_2(\Lambda(\frac{3}{2}^-)_3\Sigma)$ = 1.
Similarly, the kaon cloud contribution to the flavor singlet is
$-34$ MeV as shown in Eq.~(\ref{eq:BFOUR}) and this includes a $\bar{K} N$
intermediate state of contribution of
$\frac{10}{3}Re\delta_2(\Lambda(\frac{3}{2}^-)_3N)k^K_{SA'A'S} = -3.73$ MeV
when $Re\delta_2(\Lambda(\frac{3}{2}^-)_3N)$ = 1.
The corresponding contributions to the flavor singlet component of
$\Lambda(\frac{1}{2}^-)_3$ are as follows (see {\it e.g.}
Eq.~(2c) in Ref.~\cite{um2}).
For the pion cloud we have a contribution of $-62$ MeV which includes an
intermediate $\pi \Sigma$ state contribution of
$2Re\delta_0(\Lambda(\frac{1}{2}^-)_3\Sigma)m^{\pi}_{SPPS} = -8.7$ MeV
when $Re\delta_0(\Lambda(\frac{1}{2}^-)_3\Sigma)$ = 1, and
the kaon cloud mass contribution is $-13$
MeV of which the $\bar{K} N$ intermediate state
contribution is
$4Re\delta_0(\Lambda(\frac{1}{2}^-)_3N)m^K_{SP'P'S} = -1.85$
MeV when $Re\delta_0(\Lambda(\frac{1}{2}^-)_3N)$ = 1.
We note that when we calculate the spin-flavor matrix elements between the
various
baryon states we use the prescription of Ref.~\cite{mw}, (see
Fig.~4 of this reference) where the quark remains in its initial state
when $i = j$ in the sum over the quark indices $i$ and $j$ in
Eq.~(\ref{eq:THREE}). The consequence of this prescription is
that only the terms $i \ne j$ in Eq.~(\ref{eq:THREE}) contribute to the
$\pi \Sigma$ and $\bar{K} N$ intermediate states.
(We relax this restriction when we calculate the hadronic widths of the
hyperons in Section~2.3 of this paper.)

Assuming now that $\Lambda(\frac{3}{2}^-)_3$ and $\Lambda(\frac{1}{2}^-)_3$
are pure flavor singlets, we find, with no mass
correction factors ({\it i.e.,} with all $Re\delta_L(BB')$ =1), the following
masses.
\beqa
M_{\Lambda(\frac{3}{2}^-)_3} & = & (1695 - 104 - 34)\: {\rm MeV} \nonumber\\
                             & = & 1557\: {\rm MeV}\\
M_{\Lambda(\frac{1}{2}^-)_3} & = & (1695 - 62  - 13)\: {\rm MeV} \nonumber\\
                             & = & 1620\: {\rm MeV}
\eeqa
Here the three terms in the parenthesis
are the "bare" $\Lambda^*$ mass, the pion- and the kaon- cloud
contributions, respectively.
When we include the estimated values of the
real part of the mass correction factors quoted
above, we find the following changes to the above masses.
\beqa
M_{\Lambda(\frac{3}{2}^-)_3} & = & (1695 - 154 - 49)\: {\rm MeV}\nonumber\\
                             & = & 1492\: {\rm MeV}\\
M_{\Lambda(\frac{1}{2}^-)_3} & = & (1695 - 79  - 43)\: {\rm MeV}\nonumber\\
                             & = & 1573\: {\rm MeV}
\eeqa
Although the mass correction factors for the $\bar{K} N$ intermediate states
are all very large, the kaon cloud contributions are too small to invert the
calculated mass ordering. The reason is due to the small values
of the kaon cloud transition matrix elements, $k^K_{SA'A'S}$ and
$m^K_{SP'PS}$, a result of the heavy kaon mass. It
seems to be necessary to go beyond lowest order of this model, perhaps by
including the non-perturbative effects of the meson cloud, to
explain the mass splitting of the $\lama$ and $\lamb$. Note that in the Cloudy
Bag Model of Veit {\it et.al.}$^{\cite{cbag}}$ an additional, very attractive,
four point quark-meson interaction of the type $\bar{q}\gamma^{\mu} q
(\phi \times \partial_{\mu} \phi)$, which is second order in the meson field,
is included leading to a $L$=0 meson-baryon contact interaction.  The effect
of this interaction term on the $\lama$ state should be examined.
\vfill\eject
\setcounter{equation}{0}
\renewcommand{\theequation}{\thesection-\arabic{equation}}
\section{$Y\bar{K}N$ Coupling Constants}
In this Appendix we briefly present the calculation of the hyperon
coupling constants $G_{\lamz\bar{K}N}$ and $G_{\sigz\bar{K}N}$ in the chiral
bag
model where the meson cloud is treated as a perturbation. The method employed
in determining $G_{Y\bar{K}N}$ is a straightforward generalization of the
calculation of $\pi NN$
coupling constant as described in Ref.~\cite{myh}. The basic assumption made
here is the identification of the hyperon coupling constant $G_{Y\bar{K}N}$
with the usual $\pi NN$ coupling constant in the $SU(3)$
limit ({\it i.e.} the
$s$-quark mass equals the u- and d- quark masses {\it and}
the kaon mass becomes equal to the pion mass).

If the quark-meson coupling at the bag surface is linearized as in
Eq.~(\ref{eq:ONE}), then the $K^-$ field generated by a strange quark in the
$S$ state can be written as
\beqa
\lefteqn{K^-(\vec{r}\, ) =} \nonumber \\
& & -\frac{A(-1)N(-1)}{4\pi f_K} \frac{\mu^2 e^{\mu R}}{(\mu R
+1)^2+1} \frac{1+\mu r}{(\mu r)^2} e^{-\mu r} \left\bra \sum_i \vec{\sigma}(i)
V_+(i) \right\ket \cdot \hat{r}
\label{eq:CONE}
\eeqa
where $r \geq R$.
Here $\mu$ is the kaon mass, $f_K$ is the kaon decay constant and the quark
index $i$ runs from 1 to 3.
$V_+$ is the V-spin raising operator acting on the flavor wavefunction.
The coefficients $A(-1)$ and $N(-1)$ are
proportional to the normalization constants for the $s$-quark and $u$- or
$d$-quark wavefunctions in the bag cavity, respectively, so that in the
$SU(3)$ limit $A(-1) \rightarrow N(-1)$.
In close analogy with the calculation of $\pi NN$ coupling constant, {\it
define} the quark-kaon coupling constant $g_{qK}$ as
\beqa
\frac{g_{qK}}{M} & \equiv & \frac{A(-1)N(-1)}{f_K}
\frac{e^{\mu R}}{(\mu R + 1)^2+1}
\label{eq:CTWO}
\eeqa
so that the $K^-$ field outside the bag becomes
\beqa
K^-(\vec{r}\, ) & = & -\frac{\mu^2}{4\pi} \frac{g_{qK}}{M}
\frac{1+\mu r}{(\mu r)^2} e^{-\mu r} \left\bra \sum_i \vec{\sigma}(i)
V_+(i) \right\ket \cdot \hat{r}.
\label{eq:CTHREE}
\eeqa
In Eq.~(\ref{eq:CTWO}), $M$ is the nucleon mass and in the
$SU(3)$ limit $g_{qK}$
reproduces the quark-pion coupling constant discussed in
Ref.~\cite{myh}.

The coupling constant for the $\lamz$ and $\sigz$ hyperons are determined by
relating the expectation value of the quark operator $\sum_i \vec{\sigma}(i)
V_+(i)$ between the relevant baryon states to that evaluated at the
meson-baryon level. For example let
\beqa
\bra p\uparrow |\sum_i \vec{\sigma}(i) V_+(i)| \lamz\uparrow \ket & = & a
U_N^{\dagger}\vec{\sigma} V_+ U_{\lamz}
\label{eq:CFOUR}
\eeqa
where $\uparrow$ indicates "spin up" and $U_{\lamz}$ and $U_N$ are the Pauli
spinors for $\lamz$ and the nucleon, respectively. Then the magnitude of the
physical $\lamz\bar{K}N$ coupling constant, denoted as $G_{\lamz\bar{K}N}$ in
Section~2.3, is given by
\beqa
|G_{\lamz\bar{K}N}| & = & a|g_{qK}|
\label{eq:CFIVE}
\eeqa
A standard calculation gives
\beqa
|g_{qK}| & = & \left\{ \begin{array}{c}
                  \frac{1}{\sqrt{2}}|G_{\lamz\bar{K}N}| \\
                  \frac{3}{\sqrt{2}}|G_{\sigz\bar{K}N}|
                  \end{array}
           \right. .
\label{eq:CSIX}
\eeqa
The value of the quark-kaon coupling constant is a function of the strange
quark mass and the bag radius. With $m_S = 250$ MeV and $R = 1.125$ fm,
$|g_{qK}|$ is found to be 6.84 and the resulting magnitude of the physical
$\lamz\bar{K}N$ and $\sigz\bar{K}N$ coupling constants are 9.68 and 3.23,
respectively.
\vfill\eject
\setcounter{equation}{0}
\renewcommand{\theequation}{\thesection-\arabic{equation}}
\section{Meson Electromagnetic Currents}
In this Appendix we outline a derivation of pion electromagnetic transition
currents and
their corresponding radial integrals used to evaluate the meson cloud
contributions to the hyperon radiative
decay widths. Let $\vec{\pi}_{XY}(\vec{r})$ be the
effective pion field emitted by a quark at the bag surface in an initial state
$X$ and final state $Y$, where $X$ and $Y$ can be in any of quark states $S
\equiv S_{1/2}, P \equiv P_{1/2}$ or $A \equiv P_{3/2}$.
As discussed these pion fields are determined by requiring a continuous axial
current across the bag surface in the chiral limit, and
their derivation was presented explicitly in Appendix~A of Ref.~\cite{um1}.
The general expression for the pion field is
\beqa
\lefteqn{{\bf \pi}_{XY} \equiv \vec{\pi}_{XY}(\vec{r}\,) = } \nonumber \\
  & & \frac{i}{2f_{\pi}}\sum_{l,m}f_l(i\mu r)
\int\, d^2r' \left.\left( \bar{q}_X(\vec{r}\,') \gamma_5 \vec{\tau}
q_Y(\vec{r}\,')
\right)\right|_{r'=R} Y^*_{lm}(\hat{r}) Y_{lm}(\hat{r}')
\label{eq:AONE}
\eeqa
where
\beqa
f_l(i\mu r) & \equiv & \frac{h_l(i\mu r)}{\partial h_l(i\mu R)/\partial R}
\label{eq:ATWO}
\eeqa
\noindent and $h_l(x)$ are the spherical Hankel functions of the first kind
\beqa
h_l(x) & = & -i(-1)^l x^l \left(\frac{1}{x}\frac{d}{dx}\right)^l
\left(\frac{e^{ix}}{x}\right).
\label{eq:ATHREE}
\eeqa
Here $q_Y(\vec{r}\,)$ is the field of a quark in state $Y$, $\mu$ is the pion
mass and $Y_{lm}(\hat{r})$ is the spherical harmonic.

The pion electromagnetic transition current can easily be constructed
from the model Lagrangian given in Eq.~(\ref{eq:ONE}). Let
$\vec{J}^{ij}_{\pi}(XYZW)$ be the pion transition current emitted by a quark
$i$ in an initial state $X$ and final state $Y$ and absorbed by a quark $j$ in
an initial state $Z$ and final state $W$ (see Fig.~3a). Then
$\vec{J}^{ij}_{\pi}(XYZW)$, which in Section~3 is defined generically as
$\vec{J}_m$, can be written as
\beqa
\vec{J}^{ij}_{\pi}(XYZW) & = & -\frac{ie}{2} \left[
\left( \vec{\nabla}{\bf \pi}_{XY}^i \right)^{\dagger} {\bf \pi}_{ZW}^j -
\left( {\bf \pi}_{XY}^i \right)^{\dagger} \vec{\nabla}{\bf \pi}_{ZW}^j
\right].
\label{eq:AFOUR}
\eeqa
Since these pion fields are solutions of the free Klein-Gordon equation,
the pion electromagnetic current
$\vec{J}^{ij}_{\pi}$ is automatically conserved.
For example, using the notation of Ref.~\cite{mw},
the transition currents corresponding to the diagrams in Figs.~3b and 3c are
\beqa
\vec{J}^{ij}_{\pi}(PSSS) & = & -\frac{ie}{2} \left[
\left( \vec{\nabla}{\bf \pi}_{PS}^i \right)^{\dagger} {\bf \pi}_{SS}^j -
\left( {\bf \pi}_{PS}^i \right)^{\dagger} \vec{\nabla}{\bf \pi}_{SS}^j
\right] \nonumber \\
                         & = & +\frac{ie}{2} T_+(i)T_-(j) \left( \frac{N(S)^3
N(P)}{16\pi^2 f_{\pi}^2} \right) P^i[P \rightarrow S]  \label{eq:AFIVE} \\
                         &   & \mbox{ } \times \frac{1}{r} \left[
g_1(i\mu r) \vec{\sigma}(j)\cdot\hat{r} \vec{r}
 - f_0(i\mu r) f_1(i\mu r)\vec{\sigma}(j) \right] \nonumber
\eeqa
and
\beqa
\vec{J}^{ij}_{\pi}(ASSS) \nonumber & = & -\frac{ie}{2} \left[
\left( \vec{\nabla}{\bf \pi}_{AS}^i \right)^{\dagger} {\bf \pi}_{SS}^j -
\left( {\bf \pi}_{AS}^i \right)^{\dagger} \vec{\nabla}{\bf \pi}_{SS}^j
\right] \nonumber \\
                                   & = & +\frac{ie}{2} T_+(i)T_-(j)
\left( \frac{N(S)^3 N(A)}{8\sqrt{6}\pi^2 f_{\pi}^2} \right)
K^{[3/2,1/2]}_{ab}(i) \label{eq:ASIX} \\
                                   &   & \mbox{ } \times \frac{1}{r} \left[
f_1(i\mu r) f_2(i\mu r) \left( \vec{\sigma}(j)\hat{r}_a\hat{r}_b
- \sigma_c(j)\hat{r}_a\hat{r}_c\hat{e}_b \right) \right. \nonumber\\
                                   &   & \mbox{ }\;\;\;\;\;\;\;\;  \left.
+ g_2(i\mu r)\sigma_c(j)\hat{r}_a\hat{r}_b\hat{r}_c\hat{r}  \right]  \nonumber
\eeqa
Here the isospin operator $T_{\pm} \equiv \mp \frac{1}{\sqrt{2}}(\lambda_1
\pm i\lambda_2)$ and
$\lambda_i$ are the $3\times 3$ Gell-Mann matrices in flavor space.
The subscripts $a,b,c = 1,2,3$ indicate
the Cartesian components of the vectors or tensors and $\hat{e}$ is a
spacial unit vector. The permutation operator $P^i[X \rightarrow Y]$
permutes the states $X$ and $Y$ of quark $i$, the components of the vector
$\vec{\sigma}$ are the usual Pauli spin-matricies and
$K^{[3/2,1/2]}_{ab}$ is a quadrupole
transition operator defined in
Ref.~\cite{mw}. The function $g_n(i\mu r)$ is given by
\beqa
g_n(i\mu r) & = & f_n(i\mu r) \left( \frac{df_{n-1}(i\mu r)}{dr} -
(n-1)\frac{f_{n-1}(i\mu r)}{r} \right) \nonumber \\
            &   & \; \; \; \; - f_{n-1}(i\mu r) \left( \frac{df_n(i\mu r)}{dr}
- n\frac{f_n(i\mu r)}{r} \right).
\label{eq:ASEVEN}
\eeqa
It is evident that the transition operators resulting from these two currents
are two-body operators in quark space and no other currents need to be
constructed if one allows only ground state baryons as intermediate baryon
states.

Volume integrals needed to evaluate $\vec{I}_m$ in Eq.~(\ref{eq:SEVEN})
are straightforward but lengthy.
In the following equations,
the photon polarization vector $\hat{\epsilon}$ is expressed in Cartesian
coordinates, {\it i.e.} $\hat{\epsilon}_1 \equiv (1,0,0)$ and
$\hat{\epsilon}_2 \equiv (0,1,0)$.
For $P \rightarrow S$ transitions, the volume
integral using the pion electromagnetic current of Eq.~(\ref{eq:AFIVE}) is
\beqa
\lefteqn{ \hat{\epsilon}_l \cdot \int_R^{\infty} d^3r\,
\vec{J}^{ij}_{\pi}(PSSS) e^{-i\vec{k}\cdot\vec{r}} } \nonumber \\
& = & +\frac{ie}{2} T_+(i)T_-(j) \left( \frac{N(S)^3N(P)}{16\pi^2 f_{\pi}^2}
\right) P^{i}[P \rightarrow S] \label{eq:AEIGHT} \\
&   & \mbox{ } \times \left[ i\frac{4\pi}{3} \frac{R}{\mu} H_0(i\mu R)
H_1(i\mu R)(I_1 - I_2) \right] \sigma_l(j) \nonumber
\eeqa
Here the subscript $l$ for $\hat{\epsilon}$ and $\sigma$ is
either 1 or 2.
The volume integrals for $A \rightarrow S$ transitions involving the
current in Eq.~(\ref{eq:ASIX}) have a more complicated operator structure.
\beqa
\lefteqn{ \hat{\epsilon}_{1} \cdot \int_R^{\infty} d^3r\,
\vec{J}^{ij}_{\pi}(ASSS) e^{-i\vec{k}\cdot\vec{r}} } \nonumber \\
& = & +\frac{ie}{2} T_+(i)T_-(j) \left( \frac{N(S)^3
N(A)}{8\sqrt{6}\pi^2 f_{\pi}^2} \right) \label{eq:ANINE} \\
&   & \mbox{ } \times \left( i4\pi \frac{R}{\mu} H_1(i\mu R)
H_2(i\mu R) \right)
\left[ \left( A(\mu R)K^{[3/2,1/2]}_{11}(i) \right. \right. \nonumber \\
&   & \mbox{ } \;\;\; \left. + B(\mu R)K^{[3/2,1/2]}_{22}(i)
+ C(\mu R)K^{[3/2,1/2]}_{33}(i) \right) \sigma_1(j) \nonumber \\
&   & \mbox{ } \;\;\;\;\;\; \left. + D(\mu R)K^{[3/2,1/2]}_{12}(i) \sigma_2(j)
+ E(\mu R)K^{[3/2,1/2]}_{13}(i) \sigma_3(j) \right] \nonumber
\eeqa
and
\beqa
\lefteqn{ \hat{\epsilon}_{2} \cdot \int_R^{\infty} d^3r\,
\vec{J}^{ij}_{\pi}(ASSS) e^{-i\vec{k}\cdot\vec{r}} } \nonumber \\
& = & +\frac{ie}{2} T_+(i)T_-(j) \left( \frac{N(S)^3
N(A)}{8\sqrt{6}\pi^2 f_{\pi}^2} \right) \nonumber \\
&   & \mbox{ } \times \left( i4\pi \frac{R}{\mu}
H_1(i\mu R) H_2(i\mu R) \right)  \left[ D(\mu R)K^{[3/2,1/2]}_{12}(i)
\sigma_1(j) \right. \label{eq:ATEN} \\
&   &  \mbox{ }\;\;\;\; + \left( A(\mu R)K^{[3/2,1/2]}_{11}(i)
+ B(\mu R)K^{[3/2,1/2]}_{22}(i) \right. \nonumber \\
&   & \left.\left. \mbox{ } \;\;\;\;\;\; + C(\mu R)K^{[3/2,1/2]}_{33}(i)
\right) \sigma_2(j)
+ E(\mu R)K^{[3/2,1/2]}_{23}(i) \sigma_3(j) \right] \nonumber
\eeqa
The function $H_n(i\mu R)$ is defined as
\beqa
H_n(i\mu R) & \equiv & i\mu \left[ \frac{n}{i \mu R} h_n(i\mu R) - h_{n+1}(i
\mu R)\right]
\eeqa
and the coefficients $A(\mu R)$ through $E(\mu R)$ are given by
\beqa
A(\mu R) & = & +\frac{1}{3}I_3 - \frac{5}{21}I_4 - \frac{2}{35}I_5
- \frac{2}{35}I_6 \\
B(\mu R) & = & -\frac{1}{3}I_3 - \frac{1}{21}I_4  \\
C(\mu R) & = & +\frac{2}{3}I_3 + \frac{2}{21}I_4 + \frac{1}{7}I_5\\
D(\mu R) & = & +\frac{2}{3}I_3 - \frac{4}{21}I_4 - \frac{2}{35}I_5
+ \frac{2}{35}I_6 \\
E(\mu R) & = & -\frac{4}{3}I_3 + \frac{2}{21}I_4 + \frac{8}{35}I_5
+ \frac{2}{35}I_6
\eeqa
where,
\beqa
I_1 & = & \int_1^{\infty} dx\, j_0(kRx) \left( \frac{2}{\mu Rx}
+ \frac{1}{(\mu Rx)^2} \right) e^{-2\mu Rx} \\
I_2 & = & \int_1^{\infty} dx\, j_2(kRx) \left( \frac{1}{\mu Rx}
+ \frac{2}{(\mu Rx)^2} \right) e^{-2\mu Rx} \\
I_3 & = & \int_1^{\infty} dx\, j_2(kRx) \left( \frac{1}{\mu Rx}
+ \frac{4}{(\mu Rx)^2} \right. \nonumber \\
    &   & \hspace{3.5cm} \left. + \frac{6}{(\mu Rx)^3}
+ \frac{3}{(\mu Rx)^4}  \right) e^{-2\mu Rx} \\
I_4 & = & \int_1^{\infty} dx\, j_2(kRx) \left( \frac{1}{\mu Rx}
+ \frac{6}{(\mu Rx)^2} \right. \nonumber \\
    &   & \hspace{3.5cm} \left. + \frac{12}{(\mu Rx)^3}
+ \frac{6}{(\mu Rx)^4} \right) e^{-2\mu Rx} \\
I_5 & = & \int_1^{\infty} dx\, j_4(kRx) \left( \frac{1}{\mu Rx}
+ \frac{6}{(\mu Rx)^2} \right. \nonumber \\
    &   & \hspace{3.5cm} \left. + \frac{12}{(\mu Rx)^3}
+ \frac{6}{(\mu Rx)^4}  \right) e^{-2\mu Rx}
\eeqa
\vfill\eject
\beqa
I_6 & = & \int_1^{\infty} dx\, j_0(kRx) \left( \frac{4}{\mu Rx}
+ \frac{14}{(\mu Rx)^2} \right. \nonumber \\
    &   & \hspace{3.5cm} \left. + \frac{18}{(\mu Rx)^3}
+ \frac{9}{(\mu Rx)^4} \right) e^{-2\mu Rx}
\eeqa
Corresponding expressions for kaon electromagnetic transition currents may be
derived in a similar manner with an appropriate substitution of the isospin
operator $T_{\pm}$ by  the $V$-spin operator
$V_{\pm} \equiv \mp \frac{1}{\sqrt{2}}(\lambda_4 \pm i\lambda_5)$.
\vfill\eject
%

\vfill\eject
\baselineskip 22pt
\noindent {\bf Table 1:} Masses in MeV of low-lying negative parity hyperon
resonances calculated in the NRQM of Ref.~\cite{ik} and in the chiral bag
model in this work. The bag parameters used to calculate the entries in column
three are
$B^{1/4}$ = 145 MeV, $Z_0$ = 0.25, $\alpha_s$ = 1.5 and $m_S$ = 250 MeV.
The fourth column shows the corresponding observed states taken from
Ref.~\cite{pdg}.
\vskip 1.5in

\begin{center}
\begin{tabular}{|c|c|c|c|}          \hline
State                  & NRQM       & Chiral Bag    & Observed State
\\ \hline\hline
$\Lambda(5/2^{-})$     & 1815       & 1818          & 1810 - 1830 \\ \hline
$\Lambda(3/2^{-})_{1}$ & 1880       & 1821          & none        \\ \hline
$\Lambda(3/2^{-})_{2}$ & 1690       & 1686          & 1685 - 1695 \\ \hline
$\Lambda(3/2^{-})_{3}$ & 1490       & 1528          & 1518.5 - 1520.5 \\ \hline
$\Lambda(1/2^{-})_{1}$ & 1800       & 1698          & 1720 - 1850 \\ \hline
$\Lambda(1/2^{-})_{2}$ & 1650       & 1667          & 1660 - 1680 \\ \hline
$\Lambda(1/2^{-})_{3}$ & 1490       & 1557          & 1400 - 1410 \\ \hline
State                  & NRQM       & Chiral Bag    & Observed State
\\ \hline\hline
$\Sigma(5/2^{-})$      & 1760       & 1780          & 1770 - 1780 \\ \hline
$\Sigma(3/2^{-})_{1}$  & 1815       & 1802          & none        \\ \hline
$\Sigma(3/2^{-})_{2}$  & 1805       & 1752          & none        \\ \hline
$\Sigma(3/2^{-})_{3}$  & 1675       & 1651          & 1665 - 1685 \\ \hline
$\Sigma(1/2^{-})_{1}$  & 1810       & 1688          & 1730 - 1800 \\ \hline
$\Sigma(1/2^{-})_{2}$  & 1750       & 1653          & none        \\ \hline
$\Sigma(1/2^{-})_{3}$  & 1650       & 1645          & none        \\ \hline
\end{tabular}
\end{center}
\vfill\eject
\noindent {\bf Table 2:} Relative percentages of spin-flavor contents of
low-lying negative parity
hyperons in the NRQM of Ref.~\cite{ik} and in the chiral bag model of this
work. The composition for $\Sigma(3/2^-)_2$ state in NRQM was not given in
Ref.~\cite{ik}. In both models the $J^P=\frac{5}{2}^-$ hyperons are pure
spin-quartet, flavor-octet states.
\vskip 1.5in

\begin{center}
\begin{tabular}{|c|c|c|c||c|c|c|}          \hline
Spin-flavor state ($\rightarrow$) & \multicolumn{3}{c}{NRQM (in \%)}
         & \multicolumn{3}{|c|}{Chiral Bag  (in \%)} \\ \cline{2-7}
State ($\downarrow $)    & $|^{2}1\rangle$ & $|^{4}8\rangle$ & $|^{2}8\rangle$
  & $|^{2}1\rangle$ & $|^{4}8\rangle$ & $|^{2}8\rangle$ \\ \hline\hline
$\Lambda(3/2^{-})_{1}$ &0.2   &98.0  &1.2   &0.0   &85.7  &14.2   \\ \hline
$\Lambda(3/2^{-})_{2}$ &16.0  &1.4   &82.8  &9.8   &13.6  &76.6   \\ \hline
$\Lambda(3/2^{-})_{3}$ &82.8  &0.0   &16.0  &90.5  &0.9   &8.7    \\ \hline
$\Lambda(1/2^{-})_{1}$ &3.2   &72.3  &25.0  &3.1   &92.1  &4.7    \\ \hline
$\Lambda(1/2^{-})_{2}$ &15.2  &33.6  &56.2  &37.4  &7.7   &54.8   \\ \hline
$\Lambda(1/2^{-})_{3}$ &81.0  &0.4   &18.5  &59.5  &0.0   &40.5   \\ \hline
                                           \hline
Spin-flavor state ($\rightarrow$) & \multicolumn{3}{c}{NRQM (in \%)}
         & \multicolumn{3}{|c|}{Chiral Bag  (in \%)} \\ \cline{2-7}
State ($\downarrow $)    & $|^{2}10\rangle$ & $|^{4}8\rangle$ & $|^{2}8\rangle$
  & $|^{2}10\rangle$ & $|^{4}8\rangle$ & $|^{2}8\rangle$ \\ \hline\hline
$\Sigma(3/2^{-})_{1}$  &36.0  &57.8  &6.3   &73.9  &17.0  &9.1    \\ \hline
$\Sigma(3/2^{-})_{2}$  &      &      &      &25.7  &63.3  &11.1   \\ \hline
$\Sigma(3/2^{-})_{3}$  &6.8   &1.2   &92.2  &0.9   &19.0  &80.1   \\ \hline
$\Sigma(1/2^{-})_{1}$  &84.6  &4.4   &10.9  &91.1  &2.4   &6.5    \\ \hline
$\Sigma(1/2^{-})_{2}$  &12.3  &65.6  &21.2  &0.6   &48.7  &50.8   \\ \hline
$\Sigma(1/2^{-})_{3}$  &2.9   &29.2  &67.2  &8.3   &48.8  &43.0   \\ \hline
\end{tabular}
\end{center}
\vfill\eject
\noindent {\bf Table 3:} Hyperon radiative decay widths in keV in the
linearized approximation to the chiral bag model. In columns two and three we
show the separated incoherent parts of the
quark core ($\Gamma_q$)
and meson cloud ($\Gamma_m$) contributions to the decay width. The total
decay widths
defined in Eq.~(\ref{eq:THIRTEEN}) is shown in the fourth column. As in
Ref.~\cite{um2} we use $m_S = 250$ MeV and $R = 1.125$ fm to calculate the
widths.
\vskip 1.5in

\begin{center}
\begin{tabular}{|c|c|c|c|}          \hline
Transition & $\Gamma_{q}$  & $\Gamma_{m}$      & $\Gamma_{Total}$\\
\hline\hline
$\glla$    & 31.60         & $3\times 10^{-3}$ & 31.46           \\ \hline
$\glsa$    & 48.83         & 0.33              & 50.85           \\ \hline
$\gllb$    & 75.15         & $7\times 10^{-4}$ & 74.98           \\ \hline
$\glsb$    & 2.22          & 0.04              & 1.85            \\ \hline
\end{tabular}
\end{center}
\vfill\eject
\noindent {\large\bf Figure Captions}
\vskip 0.3in
\begin{description}
\item[Figure 1:] The mass spectrum of the negative parity $\lams$ and $\sigs$
hyperons in the chiral bag model where the meson cloud is treated as
perturbation.
In the figure we show the observed hyperon resonances with their mass
uncertainties taken from Ref.~\cite{pdg} and the calculated
masses when {\it (i)} only the pion field (solid line) and {\it (ii)} both
the pion  and the
kaon fields (dashed line) are included in the meson cloud. The vertical lines
connect the same states {\it before} and {\it after} the kaon field has been
included in the
calculation. Note that when both the pion and kaon fields are included in the
meson cloud, the mass of $\Sigma(\frac{5}{2}^-)$ state coincides with the upper
limit of the $J^P=\frac{5}{2}^-$ $\Sigma(1775)$ resonance.
\item[Figure 2:] {\it (a)} and {\it (b)}:
The quark-meson-photon contact interaction
which is {\it absent} in a chiral bag model described by the Lagrangian given
in
Eq.~(\ref{eq:ONE})
due to the {\it pseudoscalar} quark-meson coupling at the bag surface. {\it
(c)}: The only type of contribution to the meson electromagnetic transition
current
considered in this work. Here the photon (wiggly line) couples to the
meson (dashed line) in flight. {\it (d)}: Radiation of a photon from an
intermediate
baryon while the meson is in flight. Because the baryons are heavy
compared to mesons this diagram has been omitted in the
present work as well as in Ref.~\cite{cloudy}.
\item[Figure 3:] {\it (a)}: A typical diagram involving the meson
electromagnetic transition current. Here the quark $i$ (solid line)
emits a meson (dashed line) and changes its state from $X$ to $Y$. The emitted
meson radiates a photon (wiggly line) and is subsequently absorbed by the
quark $j$ which changes its state from $Z$ to $W$. {\it (b)}: The diagram
involving the pion electromagnetic transition current
$\vec{J}^{ij}_{\pi}(PSSS)$ of Eq.~(C-5). Here the dashed line represents a pion
in flight.
{\it (c)}: The corresponding diagram
involving the current $\vec{J}^{ij}_{\pi}(ASSS)$ of Eq.~(C-6).
See Appendix~C for details of their derivations.
\end{description}
\vfill\eject
%
%
%
%
%
\end{document}